\documentclass[sigconf]{acmart}

\usepackage{booktabs} 
\usepackage{algorithm, algorithmic}

\usepackage{graphicx}
\usepackage{balance}  
\usepackage{url}
\usepackage{float}
\usepackage{array}
\usepackage{caption}
\usepackage{subcaption}
\usepackage{amsmath,amssymb}
\setcopyright{rightsretained}
\newcolumntype{L}[1]{>{\raggedright\let\newline\\\arraybackslash\hspace{0pt}}m{#1}}
\newcolumntype{C}[1]{>{\centering\let\newline\\\arraybackslash\hspace{0pt}}m{#1}}
\newcolumntype{R}[1]{>{\raggedleft\let\newline\\\arraybackslash\hspace{0pt}}m{#1}}

\newcommand{\querc}{Querc}

\acmConference[CIDR'19]{Conference on Innovative Data Systems Research}{Jan 2019}{Asilomar, California USA}
\acmYear{2019}
\copyrightyear{2016}

\acmArticle{4}
\acmPrice{15.00}

\begin{document}
\title{Database-Agnostic Workload Management}
%
\author{Shrainik Jain}
\affiliation{\institution{University of Washington}}
\email{shrainik@cs.washington.edu}

\author{Jiaqi Yan}
\affiliation{\institution{Snowflake Computing}}
\email{jiaqi.yan@snowflake.net}

\author{Thierry Cruanes}
\affiliation{\institution{Snowflake Computing}}
\email{thierry.cruanes@snowflake.net}

\author{Bill Howe}
\affiliation{\institution{University of Washington}}
\email{billhowe@cs.washington.edu}

\begin{abstract}
We present a system to support generalized SQL workload analysis and management for multi-tenant and multi-database platforms.  Workload analysis applications are becoming more sophisticated to support database administration, model user behavior, audit security, and route queries, but the methods rely on specialized feature engineering, and therefore must be carefully implemented and reimplemented for each SQL dialect, database system, and application.  Meanwhile, the size and complexity of workloads are increasing as systems centralize in the cloud.  We model workload analysis and management tasks as variations on query labeling, and propose a system design that can support general query labeling routines across multiple applications and database backends.  The design relies on the use of learned vector embeddings for SQL queries as a replacement for application-specific syntactic features, reducing custom code and allowing the use of off-the-shelf machine learning algorithms for labeling.  The key hypothesis, for which we provide evidence in this paper, is that these learned features can outperform conventional feature engineering on representative machine learning tasks.  We present the design of a database-agnostic workload management and analytics service, describe potential applications, and show that separating workload representation from labeling tasks affords new capabilities and can outperform existing solutions for representative tasks, including workload sampling for index recommendation and user labeling for security audits.
\end{abstract}

\maketitle

\section{Introduction}
\label{intro}

Extracting patterns from a SQL query workload has enabled a number of important features in database systems, including workload compression~\cite{chaudhuri2002compressing}, index recommendation~\cite{chaudhuri:03}, modeling user and application behavior~\cite{tran:15,jain:16a,yu:92}, query recommendation~\cite{querie}, predicting cache performance~\cite{sapia:00,dan:95}, and designing benchmarks~\cite{yu:92}.  These techniques can be used as part of a more comprehensive approach to automate database administration~\cite{pavlo2017self}. 

However, the diversity of applications have led to a diversity of solutions, each relying on specialized feature engineering.  For example, workload summarization for index recommendation uses the structure of join and group by operators as features \cite{chaudhuri2002compressing}, query recommendation may pre-process a query into fragments before making recommendations \cite{nodirasnipsuggest}, and security audits may require user-defined functions to enforce particular policies \cite{prasang}.

In fact, the features and the algorithms to extract them tend to be the significant contributions in the papers in this space. But the state of the art in a variety of applications is to learn features automatically.  For instance, Natural Language Processing applications previously relied on parsing and labeling sentences as a pre-processing step, but now use learned vector representations almost exclusively \cite{goldberg2014word2vec,pennington2014glove}. This approach not only obviates the need for manual feature engineering and pre-processing, but also has the potential to significantly outperform more specialized methods.

We see three trends motivating an analogous role for generalized workload representations.  First, workload heterogeneity is increasing, making it difficult to maintain SQL parsers and feature extraction routines.  The number of SQL-like languages is increasing, with inconsistent support and syntax for even relatively common features such as outer joins.  Second, workload scale is increasing.  Cloud-hosted, multi-tenant database services including Redshift \cite{redshift}, Snowflake \cite{snowflake}, BigQuery \cite{melnik2010dremel} and more receive millions of queries daily from thousands of customers using hundreds of schemas; relying on brittle parsers (or worse, manual inspection) to identify query patterns that influence administration decisions is no longer tenable. 
Third, new use cases for centralized workload management are emerging.  For example, SQL debugging~\cite{grust:13}, database forensics~\cite{pavlou:13}, and data use management ~\cite{prasang} motivate a more automated analysis of user behavior patterns, and cloud-hosted multi-tenant systems motivate a more automated approach to query routing and resource allocation.

In this work, we propose \querc{}, a database-agnostic systems for mining and managing large-scale and heterogeneous workloads. 
We model workload management and analysis as a set of query labeling tasks. For instance, workload sampling can be reduced to labeling each query as present or absent in the sample, error prediction involves labeling each query with an error type, query routing involves labeling each query with a cluster resource to which the query should be routed, and so on.  
Because our framework depends only on the query text (along with typical metadata such as arrival timestamp and userid issuing the query), it can be used with any DBMS and any SQL dialect. In fact, as we will show, features learned with a workload against a particular schema and SQL dialect can be effective even when used with a \emph{different} schema and SQL dialect.  

The weakness of this approach is that it requires enormous amounts of data to be effective.  But as database products migrate to the cloud, service providers have access to workloads from a large number of customers, potentially even across different database products.  Since the input is just the query text, these diverse workloads can be processed as one very large dataset.  But the resulting vectors can still be used to train models to support specific applications, as we will show on two representative tasks: workload summarization for index selection and user prediction for security audits and routing.



\section{System Architecture}
\label{systemarch}

\begin{figure}[t]
\centering
\includegraphics[keepaspectratio=true, width=0.9\columnwidth]{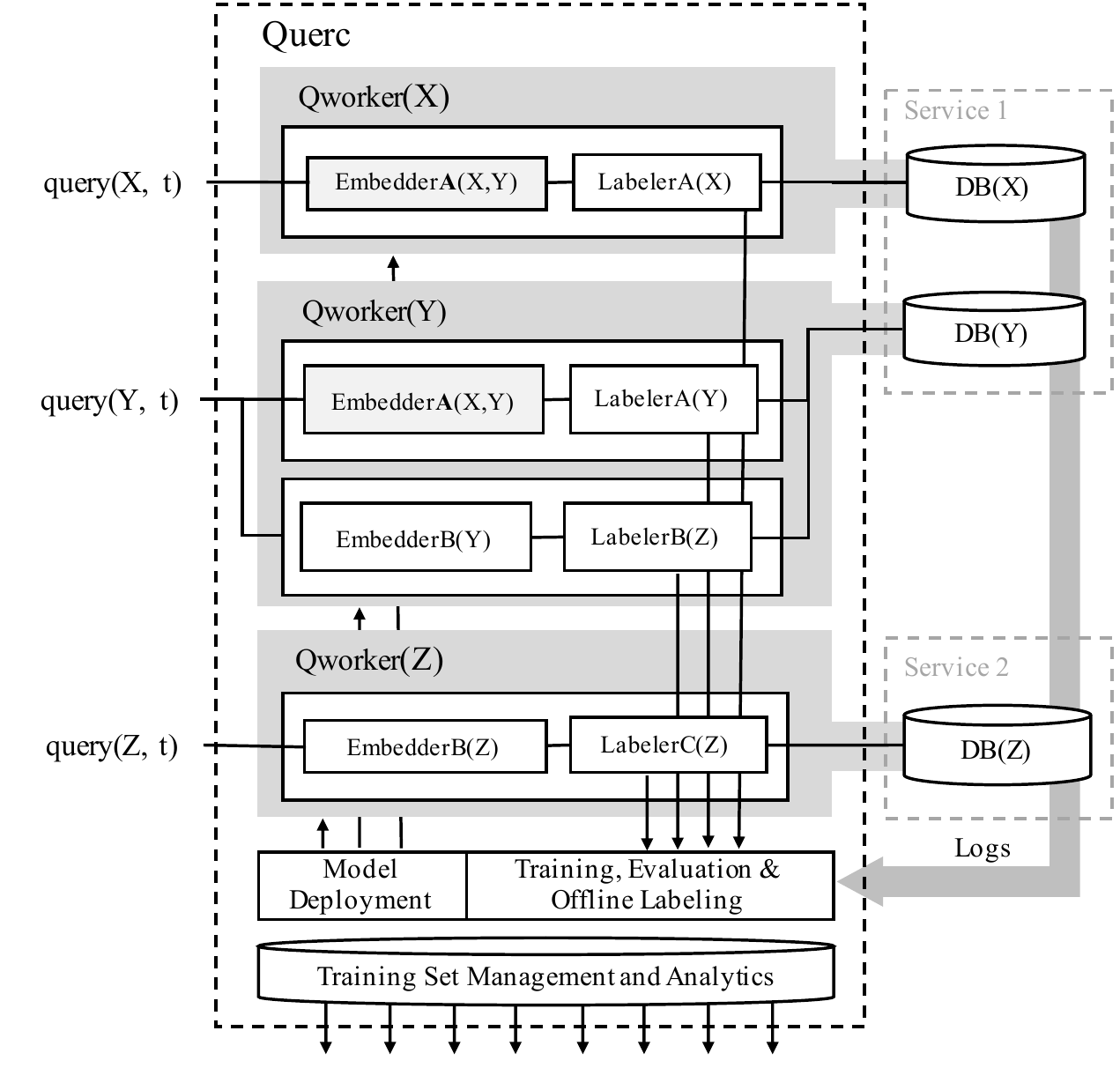}
\caption{System architecture.  Queries arrive for three different applications $X$, $Y$, and $Z$ and are processed by one or more (embedder, labeler) pair before being sent on to the database, centralized for offline labeling tasks, or both.}
\label{fig:architecture}
\vspace{-0.7cm} 
\end{figure}

Figure \ref{fig:architecture} illustrates the architecture of \querc{}.
%
%
%
%
There are three applications, X, Y, Z.  Each application has its own database, DB(X), DB(Y), and DB(Z), though these may be logical instances in the same physical multi-tenant service.  In this example, DB(X) and DB(Y) are tenants in the same service.  Each application is also associated with a separate stream of queries (at left), where query(X,t) indicates a batch of queries arriving for application X at time instant $t$.

Each application is associated with one Qworker, but each Qworker operates multiple classifiers. Qworkers may not be entirely stateless, as some labeling tasks process a small window of queries.  However, the state is assumed to be small such that the Qworkers do not need their own local storage and can be load balanced and parallelized in typical ways.  Each classifier is a pre-trained (embedder, labeler) pair.   The same trained embedder may be used across multiple applications.   This split design is critical, because we want to learn features using a very large, combined workload, but an individual classifier may perform better when trained on an application-specific workload.  In this example, application X and application Y both share the same embedder, EmbedderA, trained on the combined X and Y workloads, written EmbedderA(X,Y).  This log sharing between customers may not always be permitted by customers for security reasons, and in this example, application Z uses only its own data.  But there is some incentive for customers to pool their data as the additional signal can potentially improve accuracy, and some cloud providers support features to allow data sharing between customers.

The Labeler passes the query on to the database, but also transmits the query back to a central training module (``Training, Evaluation, and Offline Labeling'' in Figure \ref{fig:architecture}). 
The training module manages training sets, including the (parallel) execution of training and evaluation routines, then deploys trained models back to Qworkers.  There is significant ongoing research in the database, systems, and ML communities on runtime architectures for training and deploying models (e.g., \cite{li:14}); we do not discuss them further since our requirements are relatively modest.  

Since \querc{} is specialized for query workload analytics rather than general machine learning, one data model can be shared among most applications. The only messages passed between components are labeled queries.  A labeled query is a tuple $(Q, c_1,c_2,c_3,\dots)$ where $c_i$ is a label.  This simple model captures situations where a query arrives already equipped with a timestamp, a userid, an IP address, etc., but also captures more verbose query logs that are returned from the database. 

The training module also records the queries with their predicted labels for retraining, evaluation, and to support offline analysis tasks. Offline tasks are those that do not require or do not allow processing each query separately, and can be implemented as typical batch jobs. For example, query clustering  is important for workload summarization~\cite{kolaczkowski2008compressing}, but does not require real-time labeling of individual queries.  

Training data is collected periodically from the databases in the form of query logs. These logs are (batched) sequences of labeled queries, but with additional labels to be used for training, such as runtime, memory usage, error codes, security flags, resource IDs.  We do not specify the mechanism by which these logs are transmitted from the database to \querc{}, since most systems have robust means of exporting logs in appropriate forms.

In some applications, \querc{} may not be in the critical path for query execution to avoid any performance overhead or reduce dependencies. In these cases, queries will be forked to \querc{}.  No change to the architecture is required in this case; queries come in, and labeled queries are collected in the training module. The query is simply not forwarded to the database.

This architecture is not designed for continuous learning, as the training is handled separately from real time query labeling.  Not all algorithms can support fully continuous learning, and an important design goal is to support simple machine learning algorithms as labelers.  Model training is therefore assumed to occur infrequently as a batch job.  

\section{Learning Vector Representations}
\label{method}
There are multiple choices for embedders; we describe two initial models we evaluate in this paper:

\textbf{Context prediction models: } Mikolov et al. \cite{mikolov2013distributed,mikolov2013efficient,le2014distributed} proposed learning a vector representation for words by predicting the next word in a context, and then deriving a vector representation for larger semantic units (sentences, paragraphs, documents) by adding a vector representing the paragraph to each context as an additional ``word."  The learned vector for this virtual context word is used as a representation for the entire paragraph. This "Doc2Vec" method has been shown to capture semantic relationships that work well for, say, sentiment classification and clustering tasks \cite{convnetNNsentenceClassification, lecunmnist}. 
This approach can be applied directly for learning representations of SQL queries: We can use fixed-size context windows to learn a representation for each token in the query, and include an identifier to learn a representation of entire query. 

\begin{figure}
  \centering
  \includegraphics[keepaspectratio=true, width=0.8\columnwidth]{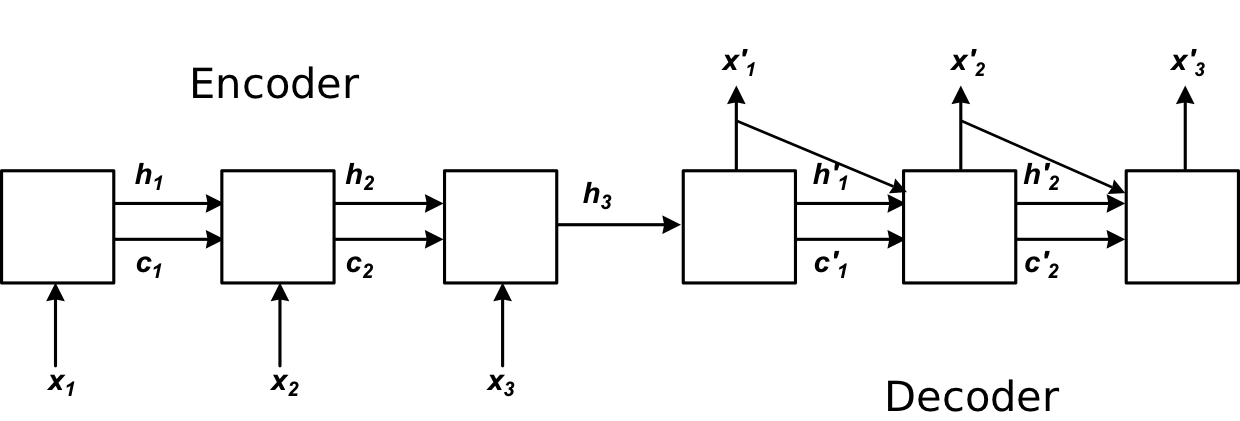}
  \caption{The LSTM Autoencoder network architecture learns to generate the input token in the decoding phase. Once trained, the encoder can be used to output a vector representation for the text of a query. }
  \label{seq_lstm_fig}
  \vspace{-1em}
\end{figure}

\textbf{LSTM AutoEncoders: } The paragraph vector approach in the previous section is viable, but it requires a hyper-parameter for the context size. There is no obvious way to determine a context size for queries, for two reasons: First, there may be semantic relationships between distant tokens in the query. Second, the length of queries vary widely in ad hoc workloads \cite{shrjainSQLShare,sqlshare_data}.
To avoid setting a context size, we can use Long Short-Term Memory (LSTM) networks \cite{zaremba2014recurrent}, which are modified Recurrent Neural Networks (RNN) that can automatically learn how much context to remember and how much of it to forget, thereby removing the dependency on a fixed context size. LSTMs have successfully been used in sentence classification, semantic similarity between sentences and sentiment analysis \cite{tang2015document}. 
We use a standard LSTM encoder decoder network \cite{auto_encoders,lstm_autoencoders} with architecture as illustrated in Figure \ref{seq_lstm_fig}.  

An LSTM autoencoder is trained by sequentially feeding words from the query to the network one word at a time, and then attempting to reproduce the input. The LSTM network not only learns the encoding for the samples, but also the relevant context window associated with the samples. The final output of the encoder network gives us an encoding for the query.
Once this network has been trained, an embedded representation for a query can be computed by passing the query to the encoder network, completing a forward pass, and using the hidden state of the final encoder LSTM cell as the learned vector representation.


There are multiple prior approaches in the NLP literature that compare the efficacy of these models and their relative performance \cite{le2014distributed, maas2011learning, tang2015document}. For this paper, we consider context-based models (i.e., doc2vec) and LSTM AutoEncoders.

\section{Applications}
The applications supported by this system reduce to \emph{query labeling}, and general workflow consists of two machine learning models: a representation learner (an embedder) and a classifier. We split the task into two parts to allow the same representation to be used for multiple applications. 

\label{applications}
\textbf{Workload summarization for index recommendation: } 
The goal \cite{chaudhuri2002compressing, kolaczkowski2008compressing} is to find a representative sample of the workload as input to further database administration, tuning, and testing tasks \cite{chaudhuri2002compressing,snowtrail}.
In particular, workload summarization aids index recommendation, since the recommendation process is typically quadratic in the size of the workload~\cite{chaudhuri2002compressing}. While index recommendation systems are well-studied and ship with most production databases \cite{chaudhuri2002compressing,chaudhuri:03}, the quality of the representative sample determines the overall quality of the final recommendations. In Section \ref{experiments}, we show that a simple sampling procedure using learned features delivers a significant runtime improvement over the built-in sampling procedure in the SQL Server database system.
%
%

\textbf{Enforcing query routing policies: } Query Routing in a distributed database involves identifying the cluster resources on which to execute the incoming query.  The policies that govern these routing decisions may involve customer SLAs, security considerations (e.g., certain applications must use a physically distinct cluster from other applications), auditing requirements (e.g., queries from certain accounts or those accessing certain tables must be logged for auditing purposes).  Even in modern cloud-hosted database products such as Snowflake \cite{snowflake} and BigQuery \cite{melnik2010dremel}, these policies tend to be manually encoded, and management of these policies as they evolve, while maintaining multiple heterogeneous clusters used by thousands of customers, is increasingly perceived as untenable.  Under the hypothesis that queries that follow a particular policy tend to have similar features, \querc{} can help identify policy misconfiguration by detecting when a predicted routing decision differs from the assigned routing decision. 
%
%

\textbf{Error prediction: } Particular syntax patterns in the workload may be associated with resource errors or bugs in the database system.  In a multi-tenant, multi-database, and high-volume scenario, identification of the syntactic patterns that tend to trigger errors, either manually or with scripts, becomes untenable: there may be hundreds of error codes, each with hundreds of subtle patterns that tend to trigger them, across hundreds of tenant schemas. Using learned features, a classifier to predict errors from syntax is trivial to engineer. This prediction allows the query to be routed to a different runtime environment that is instrumented, equipped with more more memory per node, or running a more stable version of the database engine.   We consider this application in a tech report companion to this paper \cite{query2vec}.

\textbf{Resource allocation: }
The structure of the query is not sufficient to accurately predict its runtime or memory footprint, but it can provide a hint that can be used for load balancing, scheduling, and as an input for optimization.  If we can coarsely categorize queries as memory-intensive, long-running, etc. with some degree of accuracy, these labels can be used as a simple, database-agnostic way to speculatively allocate resources.  Training data is readily available from the query logs themselves. We consider this application in a tech report companion to this paper \cite{query2vec}.
%
%

\textbf{Query recommendation: } The query recommendation problem can be modeled as a prediction of the next query  the user will submit to the database based on the recent history of queries \cite{querie}.  This prediction is then shown to the user though an appropriate client application to assist in query authoring.  Our framework can generate features that can be used to train query recommendation models.  We consider this application in a tech report companion to this paper \cite{query2vec}.

\textbf{Security auditing: } To the extent that users' individual workloads tend to follow predictable patterns, an anomalous query may be a sign that a user's account has been compromised.  By formulating a prediction problem that tries to guess the user that submitted the query from the syntax alone, we can identify anomalous queries for security audits.  In our framework, the labeler is a simple classifier $V \rightarrow user$.

\begin{figure}
\centering
\includegraphics[keepaspectratio=true, width=0.8\columnwidth]{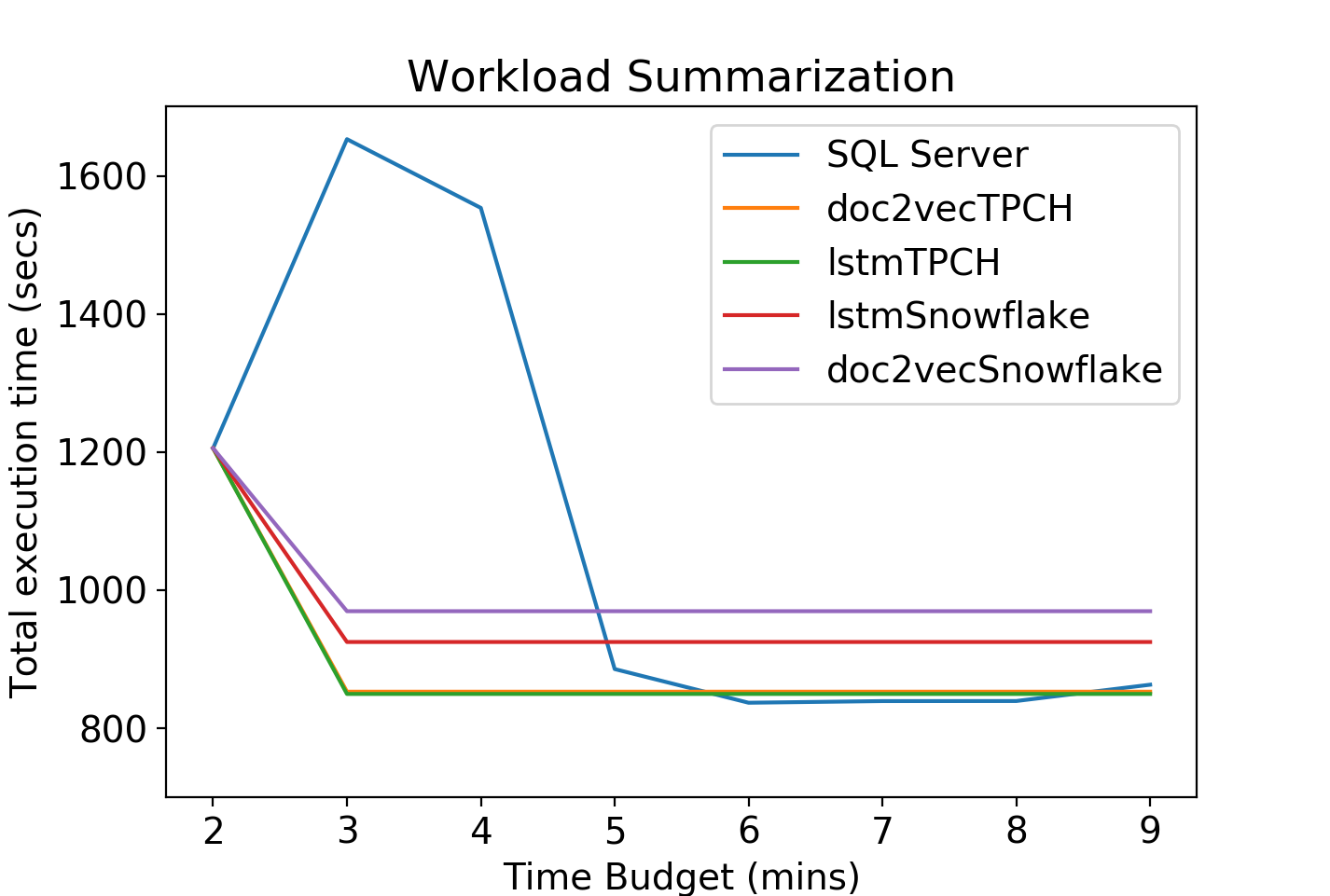}
\caption{Workload runtime using indexes recommended under various time budgets.  For most time budgets, the workload summaries improve runtimes, even when the embedders were trained on an unrelated workload (lstmSnowflake and doc2vecSnowflake).}
\label{fig:summarization}
\vspace{-1em}
\end{figure}

\section{Experiments}
\label{experiments}

We consider two applications: Workload summarization for index selection, and labeling tasks for security audits and query routing.

\subsection{Workload Summaries for Index Selection}
\label{sec_ws}

The workload summarization task (with respect to index recommendation) is to find a subset $Q_{sub}$ of a given query workload $Q$, such that the set of indexes recommended based on $Q_{sub}$ is similar to the the set of indexes recommended for the overall workload $Q$. Previous solutions are primarily variants of the approach of Chaudhuri et al.  \cite{chaudhuri2002compressing}, which uses K-medioids to cluster the queries and selects a witness query from each cluster.  However, the authors emphasize that a custom distance function should be developed for specific workloads; our hypothesis is that generic representation learning approaches obviate the need for these custom distance functions.  

In the \querc{} framework, this task is offline and does not require real-time labeling of queries. Instead, we perform the task as an offline unsupervised learning task.  In our approach, we assign each query to a vector (using a suitably trained embedder), then simply use K-means to find $K$ query clusters and pick the nearest query to the centroid in each cluster as the representative subset.  To determine $K$, we use an intentionally simple method (the ``elbow method" \cite{kodinariya2013review}) which runs the K-means algorithm in a loop with increasing $K$ till the rate of change of the sum of squared distances from centroids plateaus.  Although better methods exist, we highlight the effect of the learned vectors rather than the choice of $K$. 

\begin{figure}[t]
  \centering
  \includegraphics[width=0.8\columnwidth]{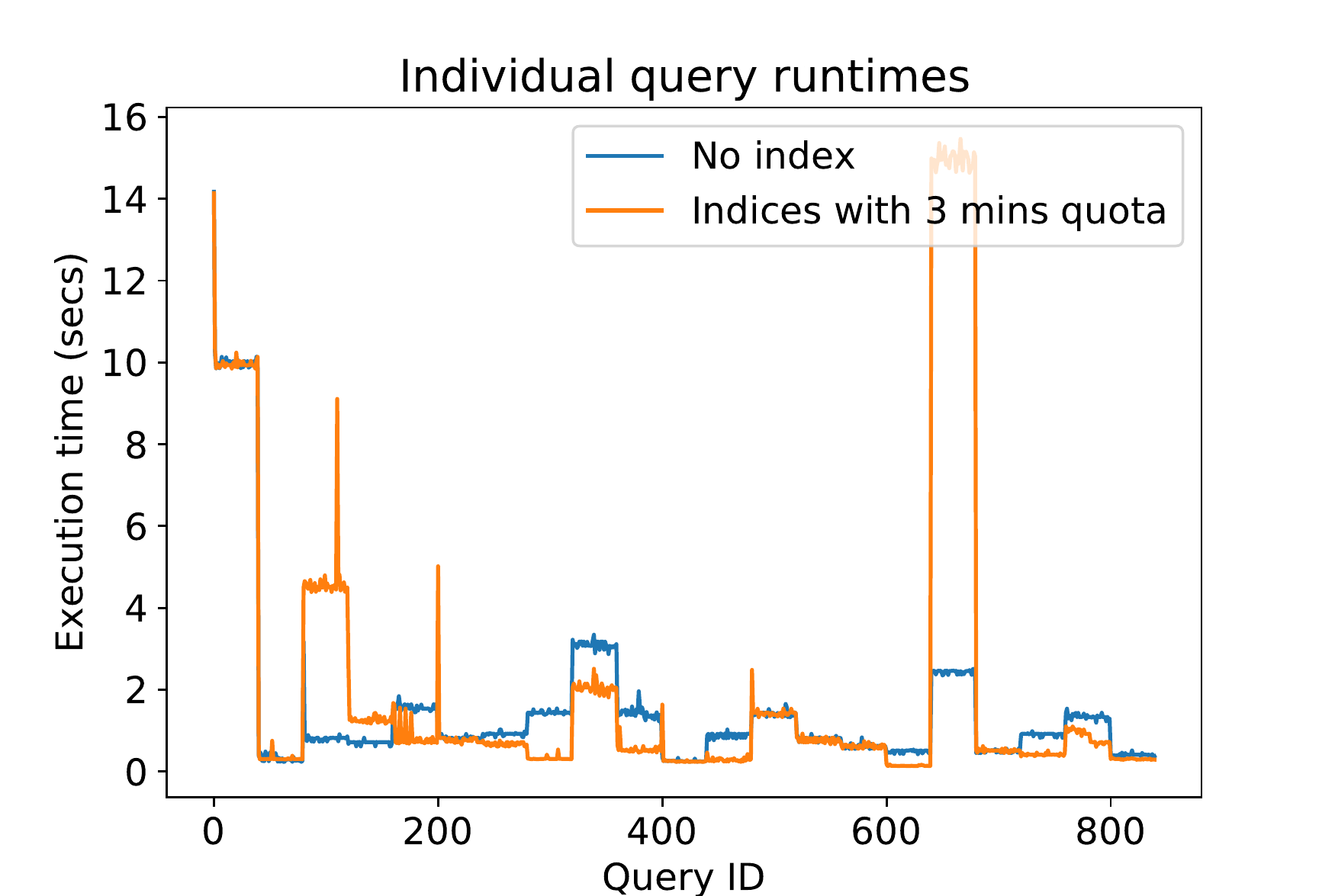}
  \caption{Runtime for each query under no indexes and under indexes recommended with a three-minute time budget.  For a few specific queries, the presence of a recommended index results in significantly worse performance.}
  \label{fig:zerovs3min}
\end{figure}

\textbf{Setup:}
Following the evaluation strategy of Chaudhuri et al.\cite{chaudhuri2002compressing}, we first run the index selection tool on the entire workload $Q$, create the recommended indexes, and measure the runtime $t_{orig}$ for the original workload. We then run use the workload summarization algorithm to produce a reduced set of queries $\mathcal{Q}_{sub}$, re-run the index selection tool, create the recommended indexes, and again measure the runtime $t_{sub}$ of the entire original workload.  
We use SQL Server 2016 and the Database Engine Tuning Advisor, which performs its own summarization on the input according to the documentation. We use an $m4.large$ AWS EC2 instance as the server.  We use TPC-H with scale factor 1 as the workload for comparison with previous results and to interpret the recommended indexes, but we also show how the method performs when trained on a more complex Snowflake workload.

We pass the summarized workload to the tuning advisor, along with a time budget (a parameter supported by the tuning advisor).  Each experiment involves clearing caches, generating indexes, applying the indexes, and running the full workload.  We report the time running the workload; the time budget specifies the time limit under which the advisor must return a set of recommendations.

\textbf{Results: } Figure \ref{fig:summarization} shows the results. The x-axis is the time budget, and the y-axis is the runtime for the entire workload after building the recommended indexes.  For time budgets less than 3 minutes, the advisor does not produce any index recommendations for any method, and the runtime is constant at 1200 seconds.  As we relax the time budget, different sets of indexes are recommended, each associated with a separate runtime.  The full workload (blue line) varies dramatically with the time budget, and surprisingly it gets worse before it gets better.  For the summarized workloads, the workload is small enough that the runtimes are constant: Once three minutes have elapsed, the advisor has found the ``optimal" set of indexes, and allowing more time does not change the result.  

We evaluate four trained embedders: two methods on two workloads.  The two methods are Doc2Vec and the LSTMAutoencoder, and the two workloads are TPC-H itself, and a separate workload of 500,000 queries from the Snowflake service.  When training the embedder on TPC-H (doc2VecTPCH and lstmTPCH), the advisor finds close-to-optimal indexes in about three minutes as opposed to the six minutes the advisor requires on the full workload.
\begin{table}[t]
\centering
\begin{tabular}{l|L{1.5cm}|L{1.5cm}|}
 & Account Labeling & User Labeling \\ \hline
Doc2Vec & 78.8\% & 39\%  \\ \hline
LSTMAutoencodder & 99.1\% &  55.4\% \\ \hline
\end{tabular}
\caption{Query Labeling results}
\label{resultsforlabeling}
\end{table}
Surprisingly, under tight time budgets, the index recommendations made by the native system can actually \emph{hurt} performance relative to having no indexes at all.  The reason is that the optimizer chooses a bad plan for a few particular queries, but the effect is enough to hurt the overall runtime.  Figure \ref{fig:zerovs3min} shows the sequence of queries in the workload on the x-axis and the runtime for each query on the y-axis under no indexes and the low-quality indexes found at the three-minute time budget.  All instances of TPC-H query 18 (queries 640-680 in Figure \ref{fig:zerovs3min}) take much longer than they would take when run without these indexes, because the optimizer finds a bad plan.  

\textbf{Transfer Learning: } Figure \ref{fig:summarization} also illustrates the capacity for transfer learning using \querc{}: When training the embedder on the snowflake dataset --- a completely unrelated workload to TPC-H workload in the \textit{SQL Server dialect }--- the summarized workload still outperforms native SQL Server for most time budgets.  This transfer learning effect allows us to bootstrap new applications without waiting for a representative workload to accumulate, and to avoid having to repeatedly re-implement brittle parsers and feature extractors for each new dialect of SQL we encounter.

\subsection{Labeling for Security Audits}
\label{sec:labeling}
We consider the conditions under which the learned features from query syntax are sufficient to predict username and customer account, where each customer has many users.  When the predicted username differs from the actual username, we can potentially flag the query for an audit.  Predicting username can help flag queries for security audits, account and cluster labels can identify misrouted queries. labels from query syntax using the two embedding methods described in Section \ref{method} over the Snowflake dataset.

\textbf{Setup: } We use embedders pre-trained on $500000$ Snowflake queries. The experiment itself is run on another dataset of $200000$ Snowflake queries labeled with username, account\_id and cluster\_name for the cluster that ran the query. Next we train classifiers (randomized decision trees) for username and customer account.

\textbf{Results: }
Table \ref{resultsforlabeling} shows the results for the labeling experiments. The numbers denote the 10-fold cross validation score on the respective task. We find that LSTM based embedders beats Doc2Vec on all tasks. The LSTM method achieves near perfect accuracy when predicting the customer account, which is because it automatically incorporates signal from the schema, and different customers use primarily different schemas (there are instances of shared schemas, but that is the less common case).  The method was completely generic and knows nothing about schemas or queries.  For user prediction, the task is more difficult, and the overall accuracy is lower at 55\%.  Upon further analysis we found that the user labeling task has $> 95\%$ accuracies for a majority of accounts (Table \ref{accountbreakdown}). The accounts that had poor accuracies for user labeling had one distinctive property: multiple users running the exact same query, making the users nearly indistinguishable.  In the sample of workload that we were working with, there were two accounts that had a number of repetitive queries by different users (for instance, $69\%$ percent of the $74000$ queries in an account had more than one user label), and these two accounts also covered around $65\%$ of the total queries, bringing down the overall accuracy of classifiers.

\begin{table}
\centering
\begin{tabular}{|c|c|c|}
\#queries & \#users & accuracy \\ \hline
73881& 28& 49.3\% \\ \hline
55333& 10& 37.4\% \\ \hline
18487& 46& 31.8\% \\ \hline
5471& 21& 96.2\% \\ \hline
4213& 6& 58.5\% \\ \hline
3894& 12& 99.7\% \\ \hline
3373& 9& 99.8\% \\ \hline
2867& 6& 99.8\% \\ \hline
1953& 15& 89.1\% \\ \hline
1924& 4& 98.1\% \\ \hline
1776& 9& 95.2\% \\ \hline
1699& 5& 99.8\% \\ \hline
1108& 12& 98.2\% \\ \hline
\end{tabular}
\caption{Top accounts with user prediction accuracy.}
\label{accountbreakdown}
\end{table}

\section{Future Work}
\label{future}
\textit{Other methods: }
There are a variety of other methods for learning representations of text that we do not evaluate in this paper. Our goal is not to identify the best possible representation learning approach but rather to show that these methods can compete with and outperform classical approaches that rely on task-specific heuristics and feature engineering (extracting JOIN clauses, counting the number of attributes, etc.), and to organize the methods into a coherent system architecture.

Alternative methods can be roughly categorized into \textit{non-neural-network} based methods and \textit{neural-network}-based methods. The non-neural-network-based methods, including non-negative matrix factorization (NMF), bag-of-words representations, and LDA \cite{maas2011learning} have been shown to be less effective than neural-network-based-methods in a variety of contexts \cite{mikolov2013distributed,levy2014linguistic}. Apart from the methods considered in this paper, there are more recent neural-network-based methods using Convolutional Neural Networks (CNNs) adapted for text data.  However, Yin et al. \cite{yin2017comparative} showed that RNN based methods (e.g., LSTMs) perform well and are robust in a broad range of tasks when compared to CNNs.
However, we plan to extend the current work to include a rigorous comparison of the techniques not covered in this paper.

\textit{Publish pre-trained models: }
The results in Section \ref{experiments} demonstrate that the proposed framework in this paper has potential to use pre-trained models on generic workloads to aid analytics for previously unseen query. In future work, we will build this framework as a service which is accessible by third parties. Given the workloads that we have access to from Snowflake \cite{snowflake}, such a service could be really beneficial for researchers who do not have access to massive query workloads.


\section{Conclusions}
\label{conclusion}
We presented the architecture for \querc{}, a database-agnostic workload analytics service that captures the structural and schema patterns present in the query workload automatically, largely eliminating the need for the specialized syntactic feature engineering that has motivated a number of papers in the literature. The proposed architecture provides a new way of organizing a variety of database administration and user productivity tasks, and provides a mechanism by which to automatically adapt database operations to specific query workloads.
Our evaluation of this architecture showed that our general framework outperformed or was competitive with previous approaches that required specialized feature engineering, and also admitted simpler classification algorithms because the inputs are numeric vectors with well-behaved algebraic properties rather than result of arbitrary user-defined functions for which few properties can be assumed. The use of transfer learning in \querc{} allows workload analytics to be \textit{SQL dialect independent} and enables the capability to bootstrap new analytics tasks and avoid re-implementing brittle codes paths.

\bibliographystyle{ACM-Reference-Format}
\bibliography{biblio}

\end{document}